\documentstyle[aps,pra,epsfig]{revtex}

\begin{document}
\draft

\title{Diffusion of two repulsive particles in a one-dimensional
lattice}

\author{Claude Aslangul\footnote
  {{\bf e-mail:} aslangul@gps.jussieu.fr}}
\address{Groupe de Physique des Solides, Laboratoire associ\'e au
CNRS (UMR 75-88), \\ Universit\'es Paris 7 \&
Paris 6, Tour 23, Place Jussieu, 75251 Paris Cedex 05, France}

\date{\today}

\maketitle
\begin{abstract}
The problem of the lattice diffusion of two particles coupled by a
contact repulsive interaction is solved by finding
analytical expressions of
the two-body probability characteristic function. The interaction induces
anomalous drift with a vanishing velocity, the average coordinate of
each particle growing at large times as $t^{1/2}$. The leading term of the
mean square dispersions displays normal diffusion, with a diffusion
constant made smaller by the interaction by the non-trivial
factor $1-1/\pi$.
Space continuous limit taken from the lattice calculations allows to
establish connection with the standard problem of diffusion of a
single fictitious particle constrained by a totally reflecting wall.
Comparison between lattice and continuous results display marked
differences for transient regimes, relevant with regards to high
time resolution experiments, and in addition show that, due to slowly
decreasing subdominant terms, lattice effects
persist even at very large times.
\end{abstract}

\vspace*{0.6cm}

\section{Introduction}\label{intro}
\setcounter{equation}{0}
Whereas classical ordinary diffusion of a single particle is
universally known, diffusion of {\it interacting} particles does not seem to
have drawn much attention.
A notable exception is the seminal paper by Fisher \cite{fischer}, which
introduces basic ideas and solves, in various cases, the problem of finding
the
probability for the reunion of a given number of ``drunken walkers''
wandering on
a one-dimensional lattice. The so-called tracer problem also received
some attention, following the solution given by Harris \cite{harris} for the
$1d$ case (recent bibliography on this subject can be found in
Mallick' thesis \cite {mallick}). More recently, intensive work
\cite{dermal}, \cite{sasawadi} and references
therein, has been done one the asymmetric simple exclusion process
(\cite{ligg}) which describes the biased motion of a lattice gas with
hard-core
interaction.

The interaction between diffusing particles appears to be relevant in many
fields: one-dimensional hopping conductivity \cite{richards}, ion
transport in biological membranes \cite{nener}, \cite{sackman},
channelling in zeolithes \cite {kukla}. Generally speaking, the
interactions are expected to play a dominant role in
low-dimensionality systems and/or sytems with geometrical constraints.
Such problems have been analyzed in the continuous space limit of the
so-called single-file model \cite{roden}, \cite{aslan}.

I here consider one of the simplest problems, namely that of two
diffusing particles on a $1d$-lattice with a repulsive contact
interaction, by directly solving the lattice master equation using
elementary methods and obtaining the exact two-body probability at all
times; this allows to find the behaviour $\forall\,t$ of the mean
square dispersion of the coordinates. The totally asymmetric
version of this problem
(directed random walk in which each particle can move only in one
direction) was treated in ref. \cite {sasawadi} using a Bethe ansatz; among
other results, these authors gave explicit asymptotic behaviour of the two
first moments of the coordinates in the case of two diffusing particles.
   
The continuous space limit of the model here considered
is simply related to the problem of a
single fictitious particle subjected to a perfectly reflecting barrier, as
shown below. On the other hand, working on a lattice seems to be usually
the most natural approach on physical grounds and is even a necessity when no
continuous limit exists; an example of such a situation is the pure
growth problem, equivalent to a directed walk, for which continuous
limit of the master equation generates a purely mechanical Liouville
equation, in which diffusive effects have disappeared \cite
{gardiner}. When the
continuous limit exists, it is expected on physical grounds that both
versions provide essentially the same results in the long time limit,
when part of
the microscopic details become irrelevant. Nevertheless, although leading
terms are expected to coincide, subdominant corrections may play a
rather important role, as shown below, since they usually follow power-laws
in time with
small exponents entailing that corrections are long-lived.
When the relevant experimental timescale is short, results obtained
by scaling hand-waving asymptotic arguments are of little physical interest
and
continuous models may even display serious shortcomings : as an example, as
shown below, the velocity at short times turns out to be
infinite in the continuous approximation whereas its lattice analogue is
perfectly well defined and is finite. In addition, somewhat surprinsingly,
lattice effects persist even in the final regime, as constrasted to
the ordinary diffusion of a single particle. This is why, as a whole,
lattice models in continuous time are worthy to investigate: the
discreteness of space, appearent in the transient as well as in the
long-time dynamics, is not a minor feature of the problem.
To be sure, lattice problems are
by nature less ``universal'' than
continuous ones in the sense that most results obtained in such a
framework usually
depend on microscopic details such as the lattice structure (\cite
{montwest}); nevertheless, they usually
contain much more physically relevant information than continuous ones and, as
such, can suggest high-resolved in time new experiments.

\section{The model}
\setcounter{equation}{0}
The basic assumptions of the present lattice model are as follows :
i) at some time ($t = 0$), a pair of particles is located on two given
adjacent lattice sites labelled $n = 0$ and $1$.
ii) when separated by more that one lattice spacing, each member of
the pair has a symmetric diffusive motion, independently of the other;
for simplicity,
it is assumed that hopping can occur between one site and its
nearest-neighbours. The hopping probability per unit time is denoted
as $W$ and allows to define a
diffusion constant $D = a^{2}W$, $a$ being the lattice spacing
and $W^{-1}$ the diffusion time.
iii) when the two particles are located on two adjacent sites,
each
of them can only move onto the empty available site: this models
the contact
interaction. As a consequence, the two particles cannot stand on
the same site and
cannot cross each other. Any initial condition thus definitely
induces a
left-right symmetry breaking. Conventionally, with the above
initial condition, the
particle located at $n = 0$ (resp. $n = 1$) at time $t=0$ will
be given the label $1$ (resp. $2$).

With these assumptions, the master equation (M. E.)
for the probability $p_{n,m}(t)$ to find the two particles
located on sites $n$ and $m$ at
time $t$ can be written following the standard procedure. One first
considers the evolution of $p_{n,m}(t)$ between $t$ and $t+\Delta t$,
where $\Delta t$ is a finite time interval, by exhausting all basic
elementary jumps and expressing their probabilities in terms of the
products $W \Delta t$; by forming the difference $p_{n,m}(t+\Delta
t)-p_{n,m}(t)$, by dividing both sides by $\Delta t$ and taking the
limit $\Delta t \rightarrow 0$, one eventually obtains the
time-continuous following M. E. ($\delta_{n,m}$ is the Kronecker symbol):
\begin{eqnarray}
	\frac{{\rm d}}{{\rm d} t}\,p_{n,m}\,& =&
	\,-\,2W\,\left[2-(\delta_{n,m+1}+\delta_{n,m-1})\right]
	(1-\delta_{n,m})\,p_{n,m} +\,W\,(p_{n+1,m}+p_{n,m-1})(1-\delta_{n,m-1})
	\,(1-\delta_{n,m})   \nonumber\\
	&+&\,W\,(p_{n-1,m}+p_{n,m+1})(1-\delta_{n,m+1})
	\,(1-\delta_{n,m})
\enspace. \label{master1d}
\end{eqnarray}
In such a way, as usual, all elementary steps involving two or more steps
disappear (their probabilities scale as $(\Delta t)^{k}$ with $k\ge 2$).
The $(1-\delta_{nm})$ factors account for the fact that the two
particles cannot stand on the same site: when $p_{nn}(0)=0$ -- which
is the case for the chosen initial condition --, then $p_{nn}(t)$
remains equal to zero at all times.

Our aim is to solve eq. (\ref{master1d}) in order to deduce the
full probability
distribution $\{p_{n,m}(t)\}$ and the related one-body (marginal, reduced)
distributions $\{p_{n}(1,t)\}$ and $\{p_{m}(2,t)\}$ given by:
\begin{equation}
	p_{n}(1,t)\,=\,\sum_{m=-\infty}^{+\infty}p_{n,m}(t)\hspace{10mm}
	p_{m}(2,t)\,=\,\sum_{n=-\infty}^{+\infty}p_{n,m}(t)
	\label{probmar}\enspace.
\end{equation}

Eq. (\ref{master1d}) can be solved by first introducing the generating
(characteristic) function $f(\phi, \psi, t)$ defined as:
\begin{equation}
	f(\phi, \psi, t)\,=\,\sum_{n,m=-\infty}^{+\infty}p_{n,m}(t)\,
	{\rm e}^{\rm i n\phi}\,{\rm e}^{\rm i m\psi}
	\label{f2carac}\enspace.
\end{equation}
allowing to find each $p_{n,m}$ by inverse Fourier transformation or to
get the marginal distributions by setting one of the
arguments $\phi$ or $\psi$ equal to zero,
and to find directly all the moments by successive derivations. It is
readily seen that the characteristic function $f$ satisfies the following
homogeneous integro-differential equation:
\begin{eqnarray}
	\frac{\partial}{\partial t}f(\phi, \psi, t)\,&=&\,
	-2W(2-\cos \phi - \cos \psi)f(\phi, \psi, t)
	+2W\,\int_{0}^{2\pi}\,\frac{{\rm d}\phi'}{2\pi}\,
	[2-\cos \phi'-\cos (\phi+\psi-\phi')]f(\phi',
	\phi+\psi-\phi', t)
	\nonumber \\
	 &+&2W\,\int_{0}^{2\pi}\,\frac{{\rm d}\phi'}{2\pi}\,
	[2-\cos(\phi-\phi')-\cos \phi-\cos \psi]
	f(\phi', \phi+\psi-\phi', t)
	\label{f2eq}\enspace.
\end{eqnarray}
The above stated initial condition writes $f(\phi, \psi, t = 0) =
{\rm e}^{\rm i\psi}$ and
entails that $p_{n,m}(t) = 0$ for $n \ge m$. By subsequently making a Laplace
transformation:
\begin{equation}
	f_{\rm L}(\phi, \psi, z)\,=\,\int_{0}^{+\infty}\,{\rm d}  t\,{\rm
e}^{-zt}
	f(\phi, \psi, t)
	\label{f2Lapl}\enspace,
\end{equation}
it is found that the function $f_{\rm L}$ obeys the following
homogeneous integral
equation:
\begin{eqnarray}
	zf_{\rm L}(\phi, \psi, z)\,-\,{\rm e}^{\rm i \psi}\,&=&\,
	-2W(2-\cos \phi - \cos \psi)f_{\rm L}(\phi, \psi, z)
	 +2W\,(2-\cos\phi-\cos \psi)\int_{0}^{2\pi}\,
	\frac{{\rm d}\phi'}{2\pi}\,f_{\rm L}(\phi', \phi+\psi-\phi', z)
	\nonumber\\
	&+&4W\,(\sin \phi-\sin\frac{\phi+\psi}{2}\,
	\cos\frac{\phi+\psi}{2})\,\int_{0}^{2\pi}\,\frac{{\rm d}\phi'}{2\pi}\,
	\sin \phi'\,f_{\rm L}(\phi', \phi+\psi-\phi', z)
	\nonumber\\
	&+&4W\,(\cos \phi-\cos^{2}\frac{\phi+\psi}{2})
	\int_{0}^{2\pi}\,\frac{{\rm d}\phi'}{2\pi}\,
	\cos \phi'\,f_{\rm L}(\phi', \phi+\psi-\phi', z)
	\enspace.
	\label{f2zeq}
\end{eqnarray}
This is a Fredholm equation with a degenerate kernel; as such, its
solution is a linear combination of the three auxiliary quantities
$M(\theta)$, $X(\theta)$ and $Y(\theta)$ ($\theta=\phi+\psi$) defined as:
\begin{eqnarray}
	M(\theta)\,&=&\,\int_{0}^{2\pi}\,\frac{{\rm d}\phi'}{2\pi}\,
	f_{\rm L}(\phi', \phi+\psi-\phi', z)\nonumber\\
	X(\theta)\,&=&\,\int_{0}^{2\pi}\,\frac{{\rm d}\phi'}{2\pi}\,
	\cos \phi'\,f_{\rm L}(\phi', \phi+\psi-\phi', z)\\ \label{MXY}
	Y(\theta)\,&=&\,\int_{0}^{2\pi}\,\frac{{\rm d}\phi'}{2\pi}\,
	\sin \phi'\,f_{\rm L}(\phi', \phi+\psi-\phi', z)\nonumber
\end{eqnarray}
$M$, $X$ and $Y$ satisfy an inhomogeneous system which can be
written down after a
somewhat lengthy but straightforward calculation:
\begin{eqnarray}
	(1-c_{11})M-c_{12}X-c_{13}Y=c_{10}\nonumber\\
	-c_{21}M+(1-c_{22})X-c_{23}Y=c_{20}\\ \label{SystMXY}
	-c_{31}M-c_{32}X+(1-c_{33})Y=c_{30}\nonumber
\end{eqnarray}
where the various coefficients are equal to:
\begin{eqnarray}
	c_{10}&=&\frac{{\rm e}^{\rm i\theta/2}\,{\rm e}^{-u}}
	{4W\cos(\theta/2)\,\sinh u}\hspace{40pt}
	c_{20}=\frac{{\rm e}^{\rm i\theta} +{\rm e}^{-2u}}
	{8W\mid\cos(\theta/2)\mid\,\sinh u}\hspace{40pt}
	c_{30}=\frac{{\rm e}^{\rm i\theta} -{\rm e}^{-2u}}
	{8\rm i W\mid\cos(\theta/2)\mid\,\sinh u}\nonumber\\
	c_{11}&=&\frac{1-{\rm e}^{-u}\mid\cos(\theta/2)\mid}
	{\mid\cos(\theta/2)\mid\,\sinh u}\hspace{40pt}
	c_{12}=\frac{{\rm e}^{-u}-\mid\cos(\theta/2)\mid}
	{\sinh u}\hspace{60pt}
	c_{13}=\frac{\tan(\theta/2)}{\sinh u}
	({\rm e}^{-u}-\mid\cos(\theta/2)\mid)\nonumber\\
	c_{21}&=&\frac{{\rm e}^{-u}}{\sinh u}(1-\cosh u\,
	\mid\cos(\theta/2)\mid)	\hspace{90pt}
	c_{22}={\rm e}^{-u}\,\frac{\sinh u+\cos^{2}(\theta/2)\,
 	({\rm e}^{-u}-\mid\cos(\theta/2)\mid)}{\mid\cos(\theta/2)
 	\mid \sinh u}\\
 	 c_{23}&=&c_{32}=\frac{{\rm e}^{-u}\sin(\theta/2)}{2\sinh u\,
	\mid\cos(\theta/2)\mid}({\rm
e}^{-u}-\mid\cos(\theta/2)\mid)\hspace{40pt}
     \nonumber \\
    c_{31}&=&\frac{{\rm e}^{-u}\tan(\theta/2)}{\sinh u}(1-\cosh u\,
	\mid\cos(\theta/2)\hspace{70pt}
	c_{33}={\rm e}^{-u}\,\frac{\sinh u+\sin^{2}(\theta/2)\,
	({\rm e}^{-u}-\mid\cos(\theta/2)\mid)}
	{\mid\cos(\theta/2)\mid\sinh u}\nonumber
\end{eqnarray}
In these equations, the quantity $u$ is defined as follows:
\begin{equation}
	\cosh u\,=\,\frac{z+4W}{4W\,\mid\cos(\theta/2)\mid}
	\label{defu}\enspace.
\end{equation}
$u$ is a function of $z$ and $\theta$, uniquely defined
by continuity from the
branch assuming real positive values when $z$ is a real
positive number.

The solution of the above system allows eventually to write
down the
Laplace transform of the generating function (\ref{f2Lapl})
as the following ($Z = z/4W$) :
\begin{eqnarray}
	 f_{\rm L}(\phi, \psi, z)\,=\,\frac{1}{4W}\,
	\frac{{\rm e}^{\rm i\psi}\,\mid\cos\frac{\phi+\psi}{2}\mid}
	{Z+1-\frac{1}{2}(\cos\phi+\cos\psi)}\,
	\frac{R(\phi+\psi,Z)\,{\rm e}^{\rm i(\phi-\psi)/2}-
	\mid\cos\frac{\phi+\psi}{2}\mid}{R(\phi+\psi,Z)-\cos^{2}\frac{\phi+\psi}{2}}
\label{fLsol}\enspace,
\end{eqnarray}
where:
\begin{equation}
	R(\theta,Z)\,\,=\,Z+1-\left[(Z+1)^{2}-\cos^{2}(\theta/2)\right]^{1/2}
	\label{Rdef}\enspace.
\end{equation}
This is a first writing of the full solution of the present problem,
since it allows to find all the moments of the two-body distribution
$\{p_{n,m}(t)\}$ in their Laplace representation. In addition, Laplace
inversion of
eq. (\ref {fLsol}) gives the characteristic function in the time domain:
\begin{eqnarray}
	f(\phi, \psi, t)\,\,&=&\,
	{\rm e}^{i\psi}\,\frac{\beta(1-\alpha){\rm e}^{i(\phi-\psi)/2}+
	\beta^{2}-\alpha}{\beta^{2}-2\alpha +1}\,{\rm e}^{4(\alpha -1)
	Wt}\nonumber\\
	&+&\,{\rm e}^{i\psi}\,(1-\beta\,{\rm e}^{i(\phi-\psi)/2})\,
	{\cal P}\int_{-\beta}^{+\beta}\frac{dx}{\pi}\frac{{\rm e}^{4Wt(x-1)}}
	{x-\alpha}\frac{(\beta^{2}-x^{2})^{1/2}}{\beta^{2}-2x+1}
	\label{fsol}\enspace.
\end{eqnarray}
In eq. (\ref{fsol}), ${\cal P}$ denotes the Cauchy principal part and :
\begin{equation}
	\alpha\,=\,\cos\frac{\phi+\psi}{2}\,\cos\frac{\phi-\psi}{2}\hspace{10mm}\beta
\,=\,\mid\cos\frac{\phi+\psi}{2}\mid
	\label{abdef}\enspace.
\end{equation}
Moreover, by setting $\psi = 0$ (resp $\phi = 0$) in eq. (\ref{fLsol}) one
readily obtains the Laplace transform of the characteristic
function for the reduced (marginal) one-body distribution for
the particle labelled $1$ (resp. $2$), $f_{1,\,\rm L}(\phi, z)$ (resp.
$f_{2,\,\rm L}(\psi, z)$);
their expressions for $- \pi \le \phi \le + \pi$ writes:
\begin{equation}
	f_{1,\,\rm L}(\phi, z)\,\frac{1}{4W}\,
	\frac{\cos\frac{\phi}{2}}{Z+\sin^{2}\frac{\phi}{2}}\,
	\frac{R(\phi,Z)\,{\rm e}^{\rm i\phi/2}-\cos\frac{\phi}{2}}
	{R(\phi,Z)-\cos^{2}\frac{\phi}{2}}
	\label{f1sol}\enspace.
\end{equation}
\begin{equation}
	f_{2,\,\rm L}(\psi, z)\,=\,{\rm e}^{\rm i \psi}
	[f_{1,\,\rm L}(\psi, z)]^{*}
	\label{f1f2}\enspace.
\end{equation}
The complex conjugation expresses the bias induced by the initial
condition and simply means that the particle $1$ tends, on the average,
to move in one direction and the particle $2$ in the opposite one.
A first derivation at $\phi = 0$ of (\ref {f1sol}) gives the
Laplace transform of
the averaged coordinate of the particle $1$, when
the other can be found anywhere else:
\begin{equation}
	x_{1,\,1}(z)\,=\,\frac{a}{4z}\,\left[1-
	\left(1+\frac{8W}{z}\right)^{1/2}\right]
	\label{m11}\enspace.
\end{equation}
A second derivation gives the second moment:
\begin{equation}
	x_{1,\,2}(z)\,=\,\frac{a^{2}}{4z}\left(1+\frac{8W}{z}\right)^{1/2}\,
	\left[\left(1+\frac{8W}{z}\right)^{1/2}-1\right]
	\label{m12}\enspace.
\end{equation}
Inverse Laplace transformation now yields the exact expressions of
the two first moments for the ``left'' particle (the one labelled $1$) :
\begin{equation}
	<x_{1}>(t)\,=\,\frac{a}{4}\,\left\{1-{\rm e}^{-T}\,
	\left[(1+2T)I_{0}(T)+2T\,I_{1}(T)\right]\right\}
	\label{m11t}\enspace.
\end{equation}
\begin{equation}
	<x_{2}>(t) = a \,- <x_{1}>(t)\label{m112t}\enspace.
\end{equation}
\begin{equation}
	<x_{1}^{2}>(t)\,=\,\frac{a^{2}}{2}\,T+\,\frac{a^{2}}{4}\,
	\left\{1-{\rm e}^{-T}\,\left[(1+2T)I_{0}(T)+2T\,I_{1}(T)\right]\right\}
	\label{m12t}\enspace.
\end{equation}
In these equations, $T$ is the dimensionless time $T = 4Wt$, whereas the
$I$'s denote the modified Bessel functions. The two mean
square dispersions coincide and can be written in the symmetric form :
\begin{equation}
	\Delta x_{1}^{2}(t)\,=\,\Delta
x_{2}^{2}(t)\,=\,2a^{2}Wt\,+\,<x_{1}>(t)<x_{2}>(t)
	\label{del2t}\enspace.
\end{equation}
It results that the mean distance between the two particles,
$d(t) = a - 2 <x_{1}>(t)$, is given by :
\begin{equation}
	d(t)\,=\,\frac{a}{2}\,
	\left\{1+{\rm e}^{-T}\,\left[(1+2T)I_{0}(T)+2T\,I_{1}(T)\right]\right\}
	\label{d(t)}\enspace.
\end{equation}
At large times ($Wt\gg 1$), the preceding expressions have the following
expansions :
\begin{equation}
	<x_{1}>(t)\,=\,-a\,\left(\frac{2}{\pi}Wt\right)^{1/2}+\frac{a}{4}+\ldots
\label{m11tinf}\enspace.
\end{equation}
\begin{equation}
	\Delta x_{1}^{2}(t)\,=\,\Delta x_{2}^{2}(t)\,=\,
	2a^{2}\left(1-\frac{1}{\pi}\right)Wt\,
	-a^{2}\,\left(\frac{Wt}{2\pi}\right)^{1/2}+
	\frac{3a^{2}}{16}+\ldots
	\label{del2tinf}\enspace.
\end{equation}
Equation (\ref {m11tinf}) shows that, due to repeated collisions with
the other
particle, the average position of one particle eventually goes to infinity,
although quite slowly ($\sim t^{1/2}$) -- and so does the average distance
$d(t)$: the repulsion induces an effective drift, characterized by a
vanishing velocity at
large times.
Equation (\ref{del2tinf}) displays the fact that the diffusion constant
of each particle is reduced by
the factor ($1 - 1/\pi$): the repulsive interaction partially inhibits the
spreading, all the more since the first correction to the linear term is
negative.

On the opposite, for short times, $t \ll W^{-1}$, one has :
\begin{equation}
	<x_{1}>(t)\,\simeq\,-a\,Wt\hspace{20mm}<x_{2}>(t)\,\simeq\,a\,(1+Wt)
	\label{m11tcourt}\enspace,
\end{equation}
\begin{equation}
	\Delta x_{1}^{2}(t)\,=\,\Delta x_{2}^{2}(t)\,\simeq\,a^{2}\,Wt
	\label{m21tcourt}\enspace.
\end{equation}

Thus, at short times, the repulsion induces a normal drift with a finite
velocity $aW$, whereas the diffusion constant is simply halved as compared
to its
value in the absence of interaction, since then each particle can
essentially diffuse in half-space only, as in a directed walk.
The two first moments
in the lattice model are plotted in Fig. \ref{fig1} (solid curves) for the
``left'' particle, using the exact expressions (\ref {m11t}), (\ref {m12t})
and (\ref {del2t}).

The above results can be compared with those given by Sasamoto and
Waditi \cite {sasawadi} for the simplified version of the
model in which each particle can move in one direction only (directed
random walk). In such a case, each particle has obviously a
finite velocity, as opposed to
the present model where jumps can occur in both directions: the
contact interaction induces for each particle a fluctuating boundary
condition which gives rise to anomalous drift, the average coordinate
increasing at all times as $t^{1/2}$. More direct comparison can be
done by considering the mean square dispersion. The quantities $<X>$
and $<X^{2}>$ introduced by these authors (their eq. (4. 11)) are
related to the above moments and are such that:
\begin{equation}
	<X^{2}>-<X>^{2}\,=\,\frac{1}{2}(\Delta x_{1}^{2}+\Delta
	x_{2}^{2})\,+\,\frac{a^{2}}{4}
	\label{Xsasa}\enspace.
\end{equation}
From (\ref{del2tinf}), it is seen that, at large times and for the
present model:
\begin{equation}
	<X^{2}>-<X>^{2}\,\simeq\,2a^{2}\left(1-\frac{1}{\pi}\right)Wt
	\label{Xsasatgd}\enspace,
\end{equation}
whereas the result of these authors for the directed walk reads:
\begin{equation}
	<X^{2}>-<X>^{2}\,\simeq\,2a^{2}\left(\frac{3}{2}-
	\frac{1}{\pi}\right)Wt
	\label{XsasaClA}\enspace.
\end{equation}
This shows that, as compared to the case without interaction, the diffusion
coefficient is enhanced for the directed walk analyzed by Sasamoto and
Waditi \cite {sasawadi} whereas it is reduced
for the both-way usual random walk model.


\section{The continuum limit}
\setcounter{equation}{0}
The continuous space limit is worthy to analyze, although it cannot
provide, by nature, the whole information contained in the lattice
version and fails to represent transient regimes; the continuous space
limit defines
an oversimplified framework and can only claim to describe features on
large space and time scales. More precisely, if $a$ is the order of
magnitude of the underlying lattice and if $\tau_{{{\rm d}iff}}$
denotes the diffusion time,
wave vectors $k$ and time-conjugate Laplace variables $z$ are physically
sensible only if they satisfy $k \ll  a^{-1}$ and
$\mid z\mid \ll \tau_{{{\rm d}iff}}^{-1}$.

\begin{figure}
\begin{center}
\epsfbox{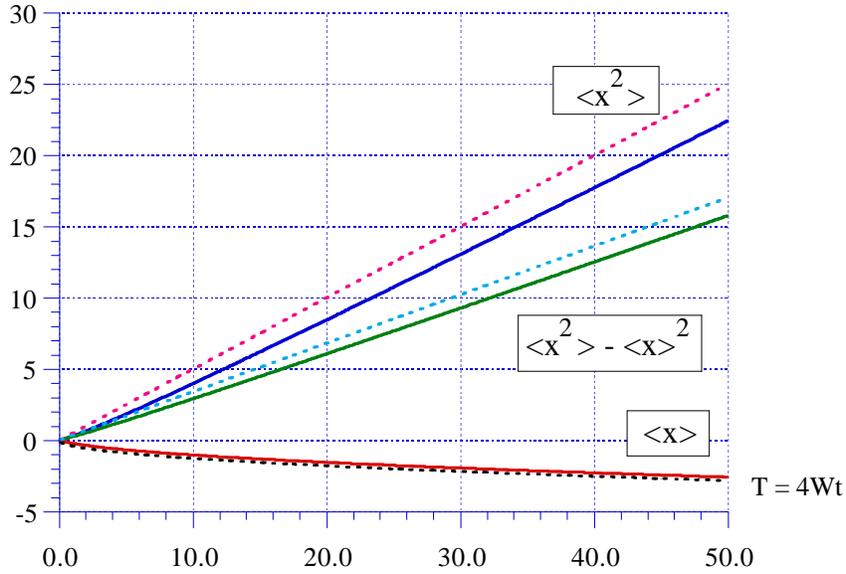}
\end{center}
\caption{Comparison of the two first moments for the reduced one-body
probability (left particle) in the lattice (solid lines, eqs. (\ref{m11t}),
(\ref{m12t}), (\ref{del2t})) and
continuous framework (dashed lines, eqs. (\ref {x1moycont}),
(\ref {deltax2cont})).}
\label{fig1}
\end{figure}

Generally speaking, the transition to continuous isotropic space is
achieved from the square lattice framework by setting $W\,=\,D/a^{2}$
and by formally taking the limit $a \rightarrow 0$. Let us set
$\phi=k_{1}a$, $\psi=k_{2}a$; when the latter limit is taken,
eq. (\ref {fsol}) generates the Laplace transform of the characteristic
function of the two-particle density, $\tilde{f_{\rm L}}(k_{1},k_{2},z)$:
\begin{equation}
	\tilde{f_{\rm L}}(k_{1},k_{2},z)\,=\,\frac{1}{\left[z+(D/2)\,
	(k_{1}+k_{2})^{2}\right]^{1/2}}
	\,\frac{1}{\left[z+(D/2)\,(k_{1}+k_{2})^{2}\right]^{1/2}\,+
	\,\rm i \sqrt{D/2}\,(k_{1}-k_{2})}
	\label{fLsolcont}\enspace.
\end{equation}
This expression is much simpler than its analogous of the discrete
version (comp. eq. (\ref {fsol})). It can be reexpressed in terms of the total
(center-of-mass momentum) $K = k_{1} + k_{2}$ and the relative (reduced)
momentum $k =  (k_{1} - k_{2})/2$ :
\begin{equation}
	\tilde{f_{\rm
	L}}(k_{1},k_{2},z)\,=\,\frac{1}{\left[z+(D/2)\,K^{2}\right]^{1/2}}\,
	\frac{1}{\left[z+(D/2)\,K^{2}\right]^{1/2}\,+\,\rm i \sqrt{2D}\,k}
	\label{fLsolcont1}\enspace.
\end{equation}
displaying the fact that the diffusion constant for the center of
mass motion is $D/2$, whereas the relative motion has a doubled
diffusion constant $2D$ (indeed $D$ plays the same role of an inverse
mass in the reduction of
a dynamical two-body problem). By using now the Efr\"{o}s theorem
\cite{laucha}, one easily
obtains:
\begin{equation}
	\tilde{f}(k_{1},k_{2},t)\,=\,{\rm e}^{-DK^{2}t/2}\,\int_{0}^{+\infty}\,
	\frac{{\rm d}\tau}{\sqrt{\pi \tau}}\,{\rm e}^{-\tau^{2}/(4t)}\,
	{\rm e}^{-\rm i \sqrt{2D}\,k\tau}
	\label{fsolcont2}\enspace.
\end{equation}
and Fourier inversion eventually yields the full two-body density
probability $P(x_{1}, x_{2}, t)$ :
\begin{equation}
	P(x_{1},x_{2},t)\,=\,\Theta(x_{2}-x_{1})\,\frac{1}{2\pi Dt}\,
	\mbox {exp}{\left[-\frac{x_{1}^{2}+x_{2}^{2}}{4Dt}\right]}
	\label{Px1x2t}\enspace.
\end{equation}
where $\Theta(x)$ is the Heaviside unit step function. This expression,
obtained as the continuum limit of expression (\ref {fsol}) is
self-evident: provided that the inequality $x_{1} < x_{2}$ is
satisfied, each particle
has a free diffusion, independently of the other.
Obviously, due to the (repulsive) interaction, $P(x_{1}, x_{2}, t)$ is not the
product of two functions, one for each particle. Indeed, $P(x_{1}, x_{2}, t)$
is an integral of such products:
\begin{equation}
	P(x_{1},x_{2},t)\,=\,\lim_{\varepsilon\rightarrow 0}\,
	\frac{1}{2\rm i\pi^{2}Dt }\,
	\int_{-\infty}^{+\infty}\,{\rm d} q\,\frac{{\rm e}^{\rm i
q(x_{2}-x_{1})}}
	{q-\rm i \varepsilon}\,{\rm e}^{-(x_{1}^{2}+x_{2}^{2})/(4Dt)}
	\label{Px1x2tFou}\enspace,
\end{equation}
exhibiting the obvious fact that the two particles are correlated at
all times.

The interparticle correlations are most simply expressed by the
correlator $C(t)$, which is readily calculated from eq. (\ref{Px1x2t}):
\begin{equation}
	C(t)\,=\,<(x_{1}-<x_{1}>)\,(x_{2}-<x_{2}>)>\,=\,<x_{1}>^{2}\,=\,
	\frac{2}{\pi}\,Dt
	\label{corrt}\enspace,
\end{equation}
showing that the normalized correlator $C(t)/<x_{i}>^{2}$ is constant
in time: the correlations induced by the interaction never die out
in one dimension.

\begin{figure}
\begin{center}
\epsfbox{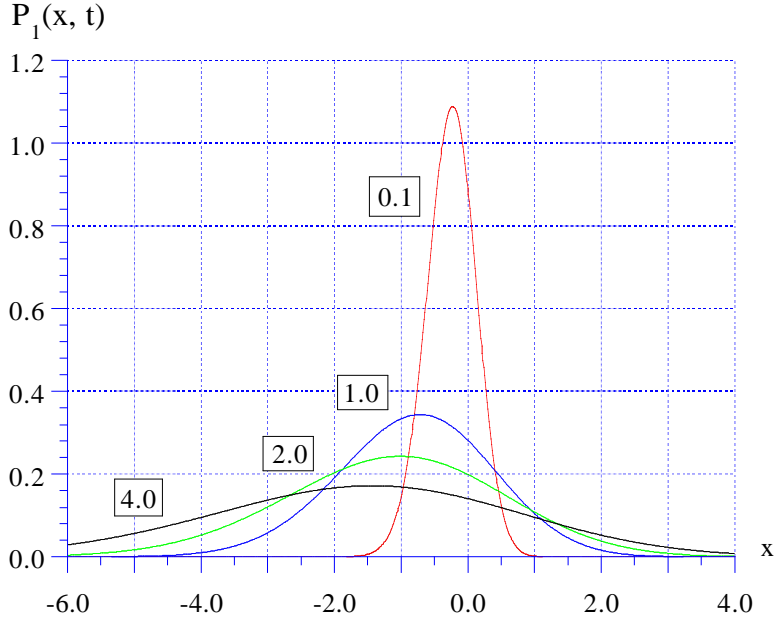}
\end{center}
\caption{Reduced one-body density for the left particle in the
continuous aproximation (see (\ref {Px1})). The diffusion constant $D$ is
taken as unity and each curve is labelled by the value of time; the
asymmetry is less and less pronounced as  time increases.}
\label{figpxt}
\end{figure}

From eq. (\ref{Px1x2t}), the marginal probabilities in space-time
for one particle, obviously asymmetric, are found to be :
\begin{equation}
	P_{1,2}(x,t)\,=\,\int_{-\infty}^{+\infty}\,{{\rm d} x'}\,P(x,x',t)\,=\,
	\frac{{\rm e}^{-x^{2}/(4Dt)}}{\sqrt{4\pi Dt}}
	\,\left[1\pm\Phi(x/\sqrt{4Dt})\right]
	\label{Px1}\enspace
\end{equation}
where the $+$ (resp $-$) sign refers to the right (resp. left)
particle and where $\Phi$ denotes the probability integral
\cite{gradry}.
The left particle marginal density is plotted in Fig. \ref {figpxt}.
Setting $k_{2} = 0$ (resp. $k_{1} = 0$) in eq. (\ref {fsolcont2}) yields
the characteristic
functions of the marginal densities:
\begin{equation}
	\tilde{f}_{1}(q,t)\,=\,{\rm e}^{-Dq^{2}t/2}\,\int_{0}^{+\infty}\,
	\frac{{\rm d}\tau}{\sqrt{\pi \tau}}\,{\rm e}^{-\tau^{2}/(4t)}\,
	{\rm e}^{-{\rm i} \sqrt{D/2}\,q\tau}
	\label{f1carc}\enspace,
\end{equation}
\begin{equation}
	\tilde{f}_{2}(q,t)\,=\,[\tilde{f}_{1}(q,t)]^{*}
	\label{f2carc}\enspace.
\end{equation}
These functions give all the moments by successive derivations at $q = 0$;
thus, at all times:
\begin{equation}
	<x_{1}>(t)\,=\,-<x_{2}>(t)\,=\,-\sqrt{\frac{2}{\pi}Dt}
	\label{x1moycont}\enspace
\end{equation}
\begin{equation}
	<x_{1}^{2}>(t)\,=\,<x_{2}^{2}>(t)\,=\,2Dt
	\label{x2moycont}\enspace
\end{equation}
\begin{equation}
	\Delta x_{1}^{2}(t)\,=\,\Delta x_{2}^{2}(t)\,=\,2(1-\frac{1}{\pi})\,Dt
	\label{deltax2cont}\enspace
\end{equation}
These expressions themselves reveal the simplifications carried out
by the continuous approximation (comp. (\ref {m11t}), (\ref {m12t})
and (\ref {del2t})); they also
display the expected fact that the exact results at all times of the
continuous version coincide with the leading asymptotic terms of the
lattice version (see (\ref {m11tinf}) and (\ref {del2tinf})); indeed,
once the dimensionless
time $T = 4Wt$ in (\ref {m11t}) and (\ref {m12t})  is replaced by
$4D/a^{2}$, the limit $a \rightarrow 0$  automatically picks out the
first leading term in the asymptotic
expansion of the Bessel functions. Note however that lattice effects
persist even at very large times: plots given in Fig. \ref{fig1} allow the
comparison between the lattice results (solid curves) and the
continuous ones (dashed curves). For the mean square dispersion, the
relative ``error'' decreases slowly as $(Wt)^{-1/2}$: it still amounts to
10\% for $Wt = 100$. The lattice effects indeed go to zero at
infinite times, but in such a slow manner that subdominant terms
cannot be neglected on reasonably large times.

On the opposite, with no surprise, the behaviours at short times are
markedly different (see (\ref {m11tcourt})
and Fig.\ref {Figxx2tc}); for instance, the finite initial drift velocity
in the lattice model
(see (\ref {m11tcourt}), which could be observed
with high-time resolution experiments, is infinite in the continuous
approximation (eq. (\ref {x1moycont})).

Note that the distribution function for the left particle (\ref{Px1})
satisfies the conservation equation:
\begin{equation}
	\frac{\partial P_{1}(x,t)}{\partial t}\,\,=\,
	- \frac{\partial J(x,t)}{\partial x}
	\label{DiffPx1}\enspace,
\end{equation}
with:
\begin{equation}
	J(x,t)\,=\,-D\,
	\left[\frac{\partial P_{1}(x,t)}{\partial
x}\,+\,2G^{2}_{D}(x,t)\,\right]
\label{CourPx1}\enspace,
\end{equation}
where $G_{D}$ denotes the normal density with a diffusion constant $D$.
The additional term to the current is negative for all $x$, as it must
clearly be.

\begin{figure}
\begin{center}
\epsfbox{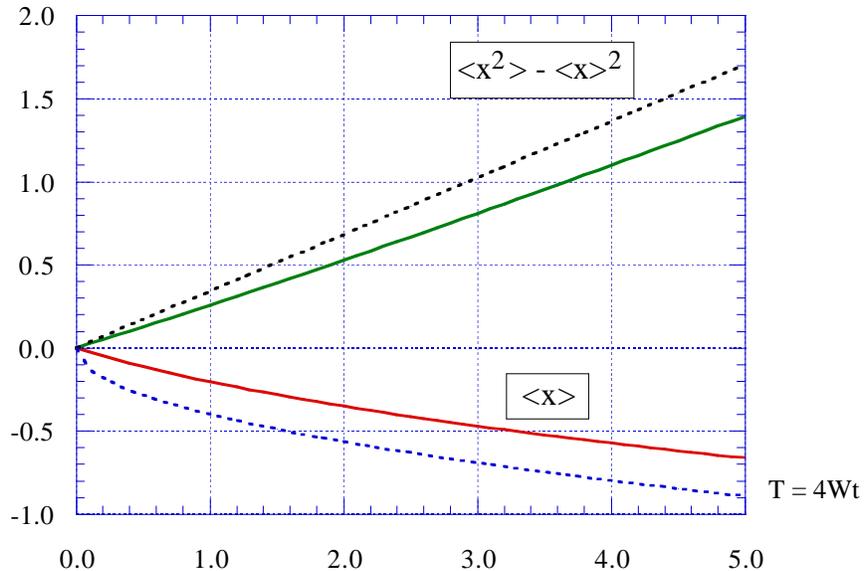}
\end{center}
\caption{Comparison at short times of the two first moments for the reduced
one-body
probability (left particle) in the lattice (solid lines), and
in the continuous framework (dashed lines).}
\label{Figxx2tc}
\end{figure}

Obviously enough, the results of the continuum approximation are
related to the simpler
problem of a single fictitious brownian particle in the presence of a
perfectly reflecting wall. For the two particles 1 and 2, the diffusion
equation writes:
\begin{equation}
	\frac{\partial}{\partial t} P(x_{1},x_{2},t)\,=\,D
	\left(\frac{\partial ^{2}}{\partial x_{1}^{2}}
	 +\frac{\partial }{\partial x_{2}^{2}}\right)\,P(x_{1},x_{2},t)
	 \hspace{5mm}\forall x_{1}\,\neq\,x_{2}
	\label{diffusion}\enspace,
\end{equation}
the solution here satisfying the initial and boundary conditions
$P(x_{1},x_{2},t=0)=\delta(x_{1}+0)\,\delta(x_{2}-0)$,
${P(x_{1},x_{2},t)=0}\hspace{2mm}{\forall x_{1}\,\ge\,x_{2}}$.
An easy way to solve such an equation is to transform to the
center-of-mass frame by setting $\quad x\,=\,x_{2}-x_{1}$,
$X\,=\,\frac{1}{2}(x_{1}+x_{2})$, $P(x_{1},x_{2},t)\,=\,\Pi(x,X,t)$;
indeed, the diffusion constant $D$ plays the same role as an inverse mass
in a mechanical two-body problem.
This allows to rewrite the diffusion equation (\ref{diffusion}) as the
following :
\begin{equation}
	\frac{\partial}{\partial t} \Pi(x,X,t)\,=\,
	\left(\frac{D}{2}\frac{\partial ^{2}}{\partial X^{2}}
	 +2D\frac{\partial}{\partial x^{2}}\right)\,\Pi(x,X,t)
	 \hspace{5mm}\forall x\,\neq\,0
	\label{diffcdm}\enspace.
\end{equation}
with $\Pi(x,X,t)=0 \quad \forall x < 0$, $\forall X\in \mbox{\bf R}$.

Equation (\ref{diffcdm}) displays the diffusion of the
center-of-mass (with a constant $D/2$), decoupled from the diffusion of the
relative
coordinate (with a constant $2D$) in the presence of a perfectly reflecting
barrier at
the origin. This means that the problem can be solved in two steps, by
considering first the diffusive motion of a single fictitious particle with
a doubled
diffusion constant and by subsequently adding the effect of the center-of-mass
motion. As a whole, the solution of the present problem in its continuous
version is given by :
\begin{equation}
	P(x_{1},x_{2},t)\,=\,G_{D/2}\left(X=\frac{1}{2}(x_{1}+x_{2}),t\right)
	\,P_{2D}(x=x_{1}-x_{2},t)
	\label{soldiff}\enspace
\end{equation}
where $G_{D/2}$ is the normalized gaussian function describing the free
diffusive
motion of the center of mass with the diffusion constant $D/2$, whereas
$P_{2D}$ is the solution for the single particle (fictitious) constrained
by the
reflecting wall. As well known (Gardiner, 1990), if the fictitious particle
stands at $x = x_{0} > 0$ with probability one at $t = 0$, one has:
\begin{equation}
	P_{2D}(x,t)\,=\,\Theta(x)\,\int_{0}^{+\infty}\,\frac{2}{\pi}\,dk\,\cos
	kx_{0}\cos kx\,{\rm e}^{-2Dk^{2}t}
	\label{barr}\enspace.
\end{equation}
By taking $x_{0} = 0+$, by using (\ref{soldiff}) and by performing all the
integrations, one recovers the expression given in (\ref{Px1x2t}),
directly obtained as
the limit taken from the lattice version.

Note however that the present problem cannot be considered as a
single particle one: the separation of the center-of-mass motion is trivial
in the sense that it happens as the consequence of discrete space homogeneity,
but this center-of-mass, itself having a diffusive motion (not a
purely kinematical one), does participate to the spreading in such a way
that correlations are always present; this is the reason why
$P(x_{1},x_{2},t)$ does not factor out (see (\ref{Px1x2t}) and (\ref
{Px1x2tFou})) and sustains never-ending correlations, as simply expressed
by (\ref {corrt}). On a deeper level, it is seen that the stochastic
motion of each particle is no more Markoffian: the Fourier
transforms (\ref{f1carc}), (\ref{f2carc}) are not of the form
${\rm e} ^{ -A\mid q\mid^{\alpha}}$ and do not satisfy the
Bachelier - Smoluchovski - Chapman -
Kolmogorov chain equation. The presence of one particle thus
strongly alters the nature of the motion of the other.


\section{Conclusions}
\setcounter{equation}{0}

The simple problem of two particles diffusing on a
lattice with contact repulsive interaction was fully solved at all times by
finding the two-particle characteristic function, from which marginal
one-particle densities and moments can be easily derived. At short times, the
repulsion induces a linear increase of the position for each particle,
associated to a finite velocity, and a linear variation of the second
cumulant,
with a diffusion constant which is halved for obvious physical reasons.
In the final regime, each particle still has an effective anomalous drift,
its coordinate growing like $t^{1/2}$, associated to a vanishing velocity.
On the other hand, the mean square dispersion asymptotically follows
a normal diffusion law; nevertheless, by comparison with its value in
the absence of interaction, the diffusion constant is reduced by the
non-trivial factor $1 - 1/\pi$. As a whole, there is a cross-over between
two normal diffusion regimes, the diffusion constant varying from $D/2$
at short times to $(1 - 1/\pi)D$ at large times. Independently of the
long-lived subdominant terms, it is seen that each particle does not
recover its ``free'' diffusion constant, even at infinite times. For
sure, this is a low-dimensionnality effect.

Taking the limit of continuous space first allows to make explicit
connection with the problem of a fictitious particle in the presence
of a totally reflecting barrier. Second, by analyzing this limit,
it can be seen that marked differences exist between the lattice
and the continuous versions. With no surprise, behaviours at short
times are very different (the initial drift is finite in the lattice,
infinite in the continuum). In addition, due to weakly decreasing
subdominant terms, behaviours still substantially differ in both
frameworks, even at large times: it can be said that
lattice effects persist for physically relevant times, formally
vanishing only at infinite times. Thus, when it is a
given physical feature, the discreteness of space cannot be ignored.


\newpage
{\bf Figure Captions}
\begin{enumerate}

\item{}
\label{fig1}
Comparison of the two first moments for the reduced one-body
probability (left particle) in the lattice (solid lines, eqs. (\ref{m11t}),
(\ref{m12t}), (\ref{del2t})) and
continuous framework (dashed lines, eqs. (\ref {x1moycont}),
(\ref {deltax2cont})).

\item{}
\label{figpxt}
Reduced one-body density for the left particle in the
continuous aproximation (see (\ref {Px1})). The diffusion constant $D$ is
taken as unity and each curve is labelled by the value of time; the
asymmetry is less and less pronounced as  time increases.

\item{}
\label{Figxx2tc}
Comparison at short times of the two first moments for the reduced one-body
probability (left particle) in the lattice (solid lines), and
in the continuous framework (dashed lines).

\end{enumerate}

\end{document}